\documentclass[aps,twocolumn,prb,10pt]{revtex4-2}
\setcitestyle{super,open={},close={}}
\usepackage{amsmath}
\usepackage{amssymb}
\usepackage[pdftex]{graphicx}
\usepackage[usenames]{color}
\usepackage{gensymb}
\definecolor{orange}{rgb}{1,0.5,0}

\begin{document}

\title {The Speed of Allosteric Signaling Within a Single-Domain Protein}
\author{ Olga Bozovic, Jeannette Ruf, Claudio Zanobini, Brankica Jankovic, David Buhrke, Philip J. M. Johnson, Peter Hamm$^*$ \\\textit{Department of Chemistry, University of
		Zurich, Zurich, Switzerland}\\  $^*$peter.hamm@chem.uzh.ch}

\begin{abstract}
\textbf{Abstract}: While much is known about different allosteric regulation mechanisms, the nature of the ``allosteric signal'', and the timescale on which it propagates, remains elusive. The PDZ3 domain from postsynaptic density-95 protein is a small protein domain with a terminal third alpha helix - the $\alpha3$-helix, which is known to be allosterically active.
By cross-linking the allosteric helix with an azobenzene moiety, we obtained a photocontrollable PDZ3 variant. Photoswitching triggers its allosteric transition, resulting in a change in binding affinity of a peptide to the remote binding pocket. Using time-resolved infrared and UV/Vis spectroscopy, we follow the allosteric signal transduction and reconstruct the timeline in which the allosteric signal propagates through the protein within 200~ns.\\


\centering{TOC Graphics:}

\centering\includegraphics[width=0.4\textwidth]{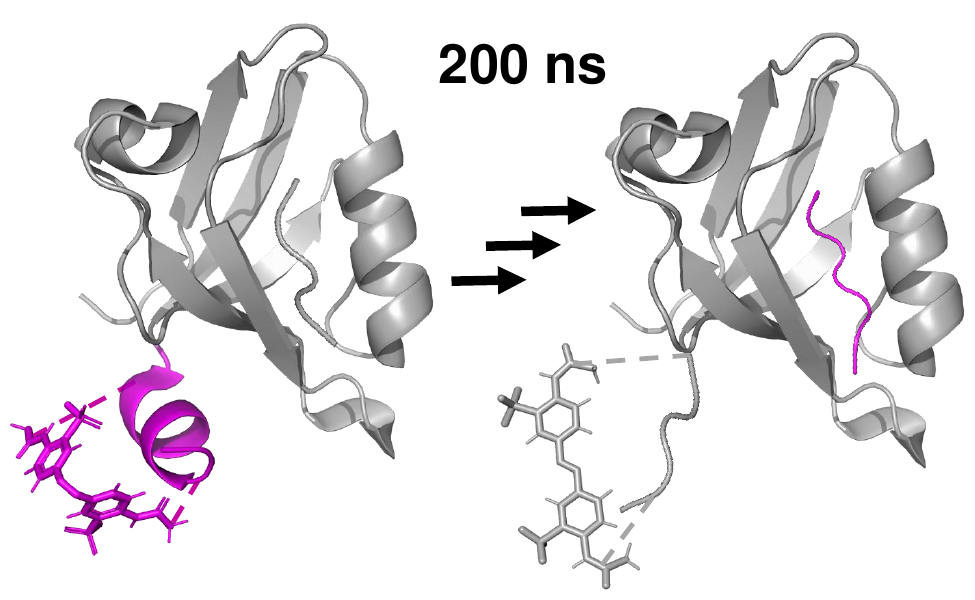}

\end{abstract}

\maketitle

To adapt is to survive, and all living cells need to constantly alter their activity according to the changing conditions of their surroundings in order to preserve homeostasis. Allosteric regulation is one of the main tools utilized by Nature to alter a target molecules' activity by adjusting its affinity to certain substrates as a response to an incoming signal. It represents the communication between two non-overlapping sites of a protein and is used as a mechanism for regulating  activity. However, the very nature of the allosteric signal is not clear.\cite{Wodak2019} Within the framework of traditional structural biology, transduction of the allosteric signal has long been considered to be the propagation of structural change throughout oligomeric proteins.\cite{monod1965,koshland1966} The concept of dynamic allostery has been introduced as well, where allostery is transduced by changing the flexibility of certain parts of a protein, and thereby its entropy, without major structural rearrangements.\cite{Cooper1984,Smock2009, petit2009,tzeng2012, saavedra2018,ahuja2019} Very recently,
long-ranged charge reorganisations have been proposed as yet another mechanism of allostery.\cite{Banerjee-Ghosh2020}


Allostery is often related to the communication between the domains of larger, multi-domain proteins.
With the discovery that isolated PDZ domains have allosteric properties as well, they became attractive systems to study allostery within single, small domain proteins.\cite{Fuentes2004, Fuentes2006,petit2009,kumawat2017,kumawat2020,gautier2018,thayer2017}  PDZ domains are a prime example, in which allostery is thought to be driven by changes in the intrinsic protein dynamics.

The PDZ3 domain from PSD-95 protein (Post Synaptic Density-95) differs from the rest of the PDZ family by an auxiliary helix at its C-terminus, and is particularly interesting in this regard.\cite{doyle1996,cabral1996,ballif2007,zhang2011,petit2009} Petit \textit{et al.} showed that even though this $\alpha$3-helix does not form any direct contact with the peptide ligand, its deletion dramatically decreases the binding affinity of the ligand to the protein by 21-fold.\cite{petit2009} In our previous work, we designed a photocontrollable variant of the PDZ3 domain, showing that allostery could be controlled by light by photoswitching the allosteric element - the $\alpha$3-helix.\cite{Bozovic2020b} To that end, an azobenzene moiety has been covalently linked to the $\alpha$3-helix in a way that photo-isomerisation of the former stabilizes/destabilizes the helicity of the latter, as an analogue of a process observed in vivo, where the phosphorylation of residue Tyr397 has the same effect.\cite{zhang2011} The construct is arguably the smallest truly allosteric protein with clearly identifiable allosteric and active sites. We were able to allosterically perturb the binding affinity for the peptide ligand by photoswitching the distal helix with astonishing changes in binding affinities, up to 120 fold depending on temperature. Conversely, the rate of the photoswitch's thermal \textit{cis}-to-\textit{trans} isomerisation is influenced by the presence or absence of the peptide ligand. By measuring binding affinities at different temperatures, we were able to construct a thermodynamic cycle of allostery and quantify the allosteric force transmitted from the peptide ligand through the PDZ3 protein to the photoswitch.\cite{Bozovic2020b}

Here, we exploit the structural sensitivity of time-resolved infrared spectroscopy,\cite{Bozovic2020a} combined with time-resolved UV/Vis spectroscopy, to follow the allosteric signal in the photoswitchable PDZ3 domain from the point of its origin – the \mbox{$\alpha$3-helix} - through the protein until it reaches the binding groove and finally transduces to the peptide.
For a spectroscopic signature specific to the source of the allosteric signal, we start with transient UV/Vis spectroscopy of the $\pi$-$\pi$*-transition of the photoswitch molecule, located on the $\alpha3$-helix. To that end, the sample was first prepared in the \textit{cis-}state by constant illumination with a 370~nm cw laser, which promotes \mbox{\textit{trans-}to\textit{-cis}} isomerisation of the photoswitch. The purity of the \textit{cis}-sample obtained in this way is estimated to be above 85$\%$.\cite{Bozovic2020a,jankovic2019} Using a short pump pulse at 420~nm, we then induce the ultrafast \textit{cis-}to-\textit{trans} isomerisation of the photoswitch, which destabilises the helical structure of the $\alpha$3-helix.\cite{Woolley2002} By following the time-resolved response of the $\pi$-$\pi$*-transition at 370~nm, we can retrieve information on the timescales of  photoswitching as well as the subsequent $\alpha$3-helix unfolding. Thus, we utilize the photoswitch molecule not only as a trigger to perturb the helical structure of the $\alpha$3-helix, and thereby initiating the allosteric response, but also as a local reporter of photoswitch conformation and $\alpha3$-helix structure.

\begin{figure}[t]
	\centering
	\includegraphics[width=0.4\textwidth]{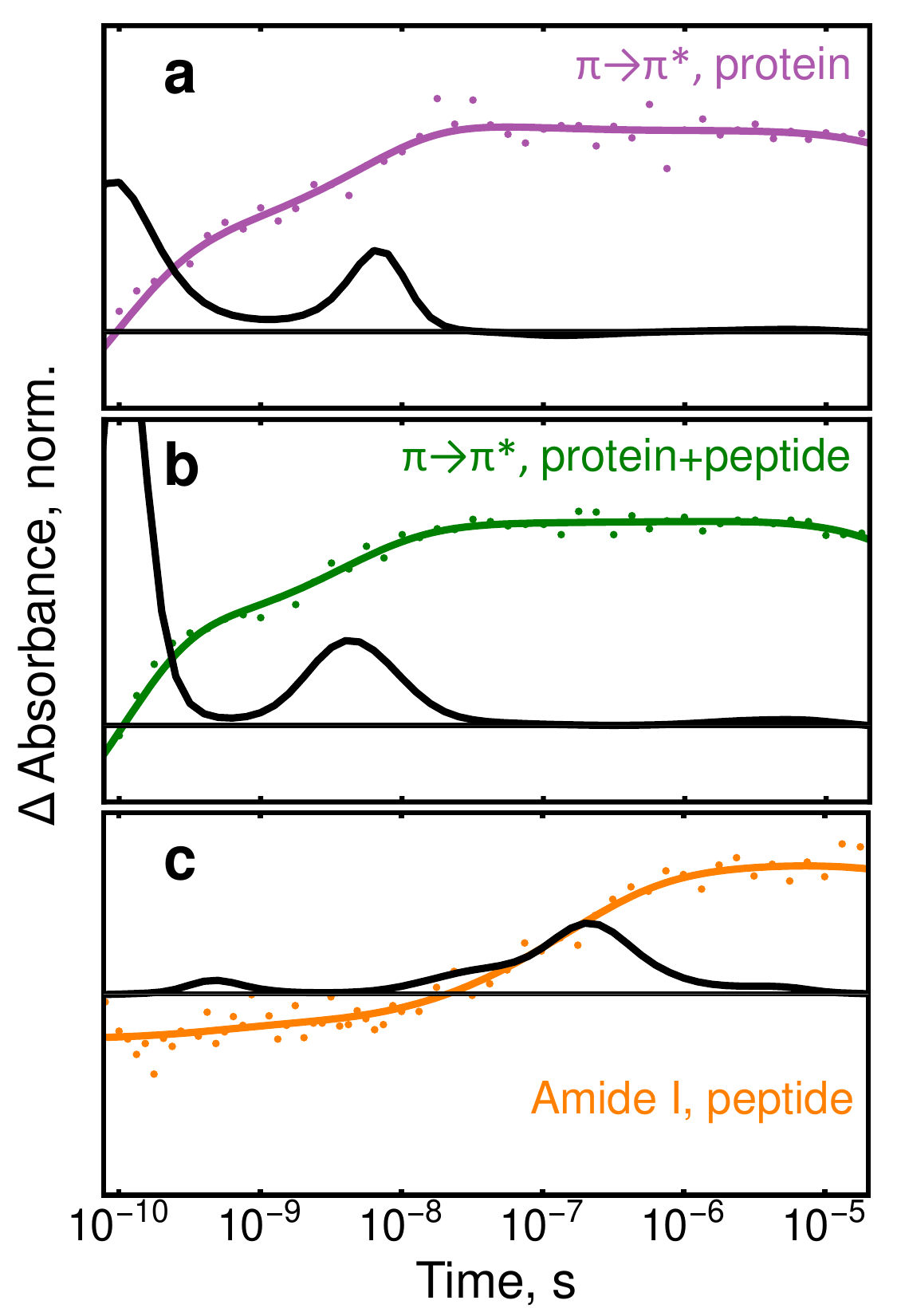}
	\caption{(a) Transient absorption changes at 370~nm for the isolated protein (purple) and (b) the protein-peptide complex (green) with the corresponding lifetime analysis (black). Panel (c) shows the equivalent for the IR double-difference response at 1630~cm$^{-1}$.}\label{uvvis}
\end{figure}

The corresponding kinetic traces are shown in Figs.~\ref{uvvis}a,b for the isolated protein and the protein-peptide complex, respectively. The timescales contained in these data are determined by fitting them to a multiexponetial function:
\begin{equation}
  S(t) = a_0- \sum_k a(\tau_{k})e^{-t/\tau_{k}} \label{EqMultiexp},
\end{equation}
where a maximum entropy method has been applied for regularisation.\cite{Lorenz-Fonfria2006,Buhrke2020}
In this fit, the timescales $\tau_{k}$ were fixed and equally distributed on a logarithmic scale with 10 terms per decade, while the amplitudes $a(\tau_k)$ were the free fitting parameters. The resulting lifetime spectra $a(\tau_k)$ are shown in Figs.~\ref{uvvis}a,b as black lines. In both cases, they reveal two similar kinetic events, one at $<$100~ps and a second one around 4~ns for the protein-peptide complex and 6~ns for the isolated protein.

The actual isomerisation of the azobenzene moiety around its central N=N bond is a barrier-less picosecond process,\cite{naeg97} giving rise to the first peak in the lifetime analysis (the time-resolution of the experiment, 60~ps, does not allow us to time-resolve this process, see Methods). Since the photoswitch is covalently linked to the \mbox{$\alpha$3-helix}, it will however be in a very constrained environment immediately following isomerisation. It has been shown that the $\pi$-$\pi$*-band is a sensitive probe of any strained conformation of the photoswitch.\cite{spo02} We therefore attribute the second kinetic component to the release of that strain, and thus to the partial unfolding of the $\alpha$3-helix. The time constant, 4 or 6~ns, respectively, is a reasonable value for the forced unfolding of an $\alpha$-helix.\cite{Ihalainen2007} The process speeds up a little bit when a peptide ligand is present in the binding pocket, as a result of the allosteric force transduced towards the photoswitch.\cite{Bozovic2020b} Unfolding of the $\alpha$3-helix represents the starting point of the allosteric signal.

\begin{figure*}[t]
	\centering
	\includegraphics[width=0.9\textwidth]{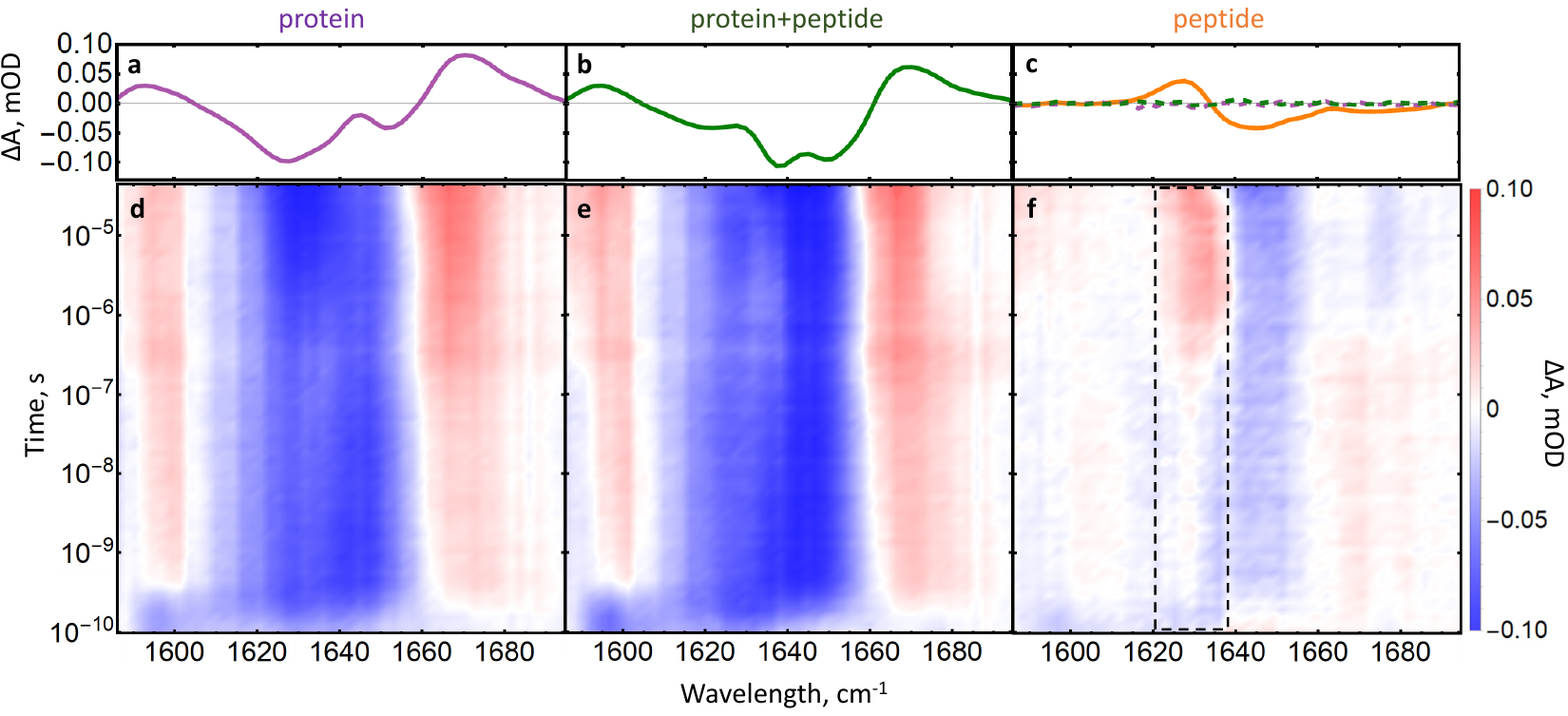}
	\caption{FTIR difference spectra for \textit{cis}-to-\textit{trans} switching of the isolated protein (a, purple), the protein-peptide complex (b, green) and double-difference spectra, revealing the response of the peptide (c, orange). In the latter case, the background noise level has been plotted as dotted lines, as determined from the data in panels (a, purple) and (b, green), see Methods for details. Panels (e-f) show the corresponding time resolved data. The dashed box in panel (f) marks the band, whose kinetics is highlighted in Fig.~\ref{uvvis}c.}
\label{IR}
\end{figure*}

To identify a spectroscopic signature of the addressee of the allosteric signal, the peptide ligand in its binding pocket, we compare FTIR difference spectra of the isolated protein (Fig.~\ref{IR}a, purple) with that of the protein-peptide complex (Fig.~\ref{IR}b, green). Each of these signals represent difference spectra upon photoswitching the samples from the \textit{cis}- to the \textit{trans}-state after effectively infinite time. There are prominent absorption changes in the amide I region between 1600 and 1700~cm$^{-1}$, arising mainly from backbone C=O modes of the protein and the peptide, which are known to be strongly structure dependent.\cite{barth02} When analyzing the response of the isolated protein, the amide~I band represents the conformational rearrangements of the protein upon photoswitching the \mbox{$\alpha$3-helix}. In contrast, in the protein-peptide complex, this band is the combined response of the protein and the peptide. Assuming that the spectra are additive, one can subtract out the protein's contribution and isolate the response of the peptide, potentially together with parts from the binding pocket that are directly affected, by taking the difference between these two spectra. Indeed, this double-difference spectrum reveals a clear spectroscopic signature (Fig.~\ref{IR}c,orange).

Figs.~\ref{IR}d,e represent the transient signals of the isolated protein and the protein-peptide complex, respectively. The two maps do not differ significantly, as the overall signal is dominated by the response of the protein. Nevertheless, by taking the difference between the two maps, we can isolate the response of the peptide (Fig.~\ref{IR}f). Fig.~\ref{uvvis}c shows the time trace of the positive band at 1630~cm$^{-1}$ together with its lifetime analysis. The first peak related to the \textit{cis}-to-\textit{trans} isomerisation of the photoswitch is largely suppressed by the subtraction, while the second peak is delayed relative to $\alpha$3-helix unfolding. We attribute its timescale, 200~ns, to the time it takes for the allosteric signal to reach the peptide ligand in its binding pocket. In contrast, the negative band at 1645~cm$^{-1}$ prevails for all times.

\begin{figure}[t]
	\centering
	\includegraphics[width=0.4\textwidth]{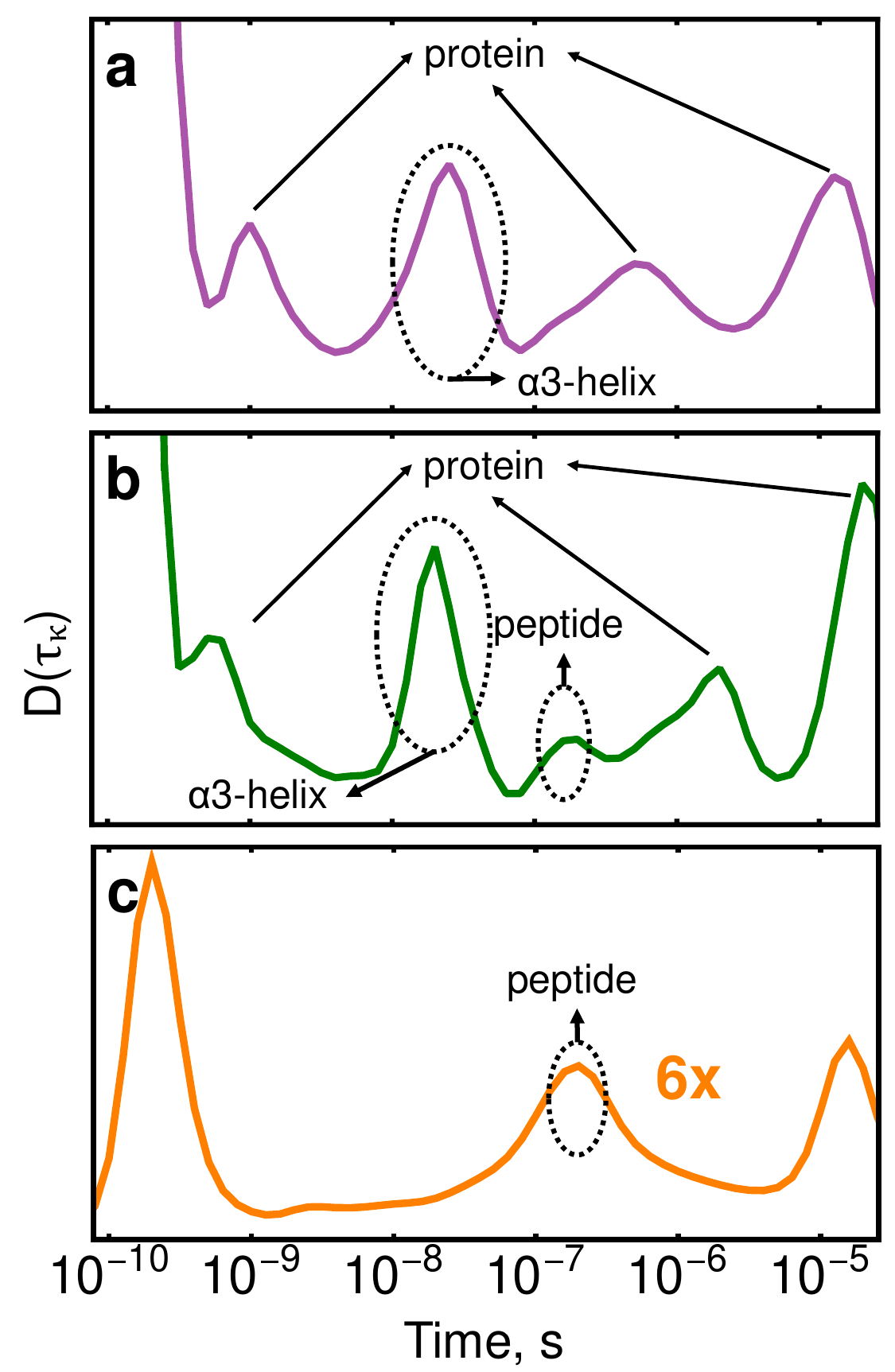}
	\caption{Dynamical content of (a) the isolated protein, (b) the protein-peptide complex and (c) the peptide ligand in its binding pocket, calculated from Eq.~\ref{Eqdyncontent}. The various timescale peaks are labeled according to their major origin.}\label{dyn}
\end{figure}

The averaged ``dynamical content'' shown in Fig.~\ref{dyn} encompasses all kinetic processes of the system.\cite{HammStock2018} It is calculated by fitting the transient data of Figs.~\ref{IR}d-f separately for each probe frequency $\omega_i$ according to Eq.~\ref{EqMultiexp}, and subsequently summing over all probe wavelengths:
\begin{equation}
  D(\tau_{k})=\sqrt{\sum_i a(\omega_i,\tau_{k})^2} \label{Eqdyncontent}.
\end{equation}
In contrast to Fig.~\ref{uvvis}c, which exemplifies the response at a single probe frequency, this expression provides an overview of all kinetic processes, and furthermore accounts for the fact that the lifetime spectra $a(\omega_i,\tau_{k})$ can have positive or negative signs.

It can be seen in Figs.~\ref{dyn}a,b that the protein responds on multiple timescales. On the other hand, when isolating the response of the peptide ligand in its binding pocket by taking a double-difference spectrum, a peak at 200~ns sticks out (Fig.~\ref{dyn}c). The dynamical content contains two additional peaks at $\approx$200~ps and 20~$\mu$s, which are significantly smaller than those in Figs.~\ref{dyn}a,b (note that the dynamical content in Figs.~\ref{dyn}c has been scaled up by a factor 6), but do not disappear completely. They do not subtract out perfectly and hence, a small contribution of the corresponding processes is also felt by the ligand and its binding pocket. We assume that this is also the reason for the prevailing negative band at 1645~cm$^{-1}$ (Fig.~\ref{IR}f). It is nevertheless obvious that the dynamical content clears up significantly, when taking the double-difference spectrum (compare Figs.~\ref{dyn}a,b vs Fig.~\ref{dyn}c).

The discrete timescales in the protein response, ranging from 1~ns to 10~$\mu$s (Figs.~\ref{dyn}a,b), very much resembles what we have observed previously for a PDZ2 domain with a photo-switchable peptide ligand. In Ref.~\onlinecite{Bozovic2020a}, we have attributed this behavior to a redistribution of population between a set of conformational substates, e.g., the local minima on the rugged energy landscape of the protein, along the lines of a Markov State Model.\cite{Bowman2013,Sengupta18}
One can think of the Markov State Model as a network of inter-converting states, whose populations shifts upon the allosteric perturbation, in accordance with one of the more recent views of allostery.\cite{Swain2006,Tsai2008,Smock2009,Tsai2014,Hilser2012} Such a network picture would not necessarily result in a ``sequence of events''. Nevertheless, when isolating the response of the peptide ligand in its binding pocket, such a sequence of events is revealed. That is, the $\alpha$3-helix unfolds on a 4~ns timescale (Fig.~\ref{uvvis}a,b), and it takes 200~ns until that signal arrives at the binding pocket (Fig.~\ref{uvvis}c and \ref{dyn}c). The more complicated response of the protein tends to mask this process; in fact the 200~ns peak in the dynamical content of the peptide ligand in its binding pocket (Fig.~\ref{dyn}c) is only a minor contribution to that of the peptide/protein complex (Fig.~\ref{dyn}b). The 200~ns timescale establishes the speed of the allosteric signal through the PDZ3 domain.

It is important to stress that the appearance of the allosteric signal at the binding pocket after 200~ns  does not yet result in unbinding of the peptide ligand. First, the off-rate, which can be estimated from the binding affinity and typical on-rates for the PDZ3 system,\cite{Gianni2005} is expected to be by many orders of magnitudes slower, 200~ms. Second, in order to distill out the structural response of the peptide ligand in its binding pocket, we have performed these experiments at room temperature, where the binding affinities in the two states of the photoswitch are essentially the same (i.e., $K_d$=3.8~$\mu$M in the \textit{trans}-state and 7.0~$\mu$M in the \textit{cis}-state at 21$^\circ$C).\cite{Bozovic2020b} Hence, the amount of bound molecules will not change, despite the very different structural and dynamical properties of the ligand in the binding pocket, which are evidenced by the fact that the ligand binding is exothermic in the \textit{cis}-state and endothermic in the \textit{trans}-state.\cite{Bozovic2020b}
What we observe with the 200~ns process is exactly this change of the structural and dynamical properties of the peptide ligand and its binding pocket.


A sequential propagation of the allosteric signal is somewhat naively thought to be only possible in large multidomain proteins, which undergo a significant conformational rearrangement. As is the example with haemoglobin\cite{Shibayama2020} or ATP synthase,\cite{boyer1997} it is natural to picture a large conformational change, which  triggers a cascade of events that finally lead to an allosteric effect. However, when discussing allosteric propagation through small protein domains, and especially ones where allostery is predominantly dynamically driven, it becomes unclear what the underlying molecular mechanism is, and what the signal sequence is.

By incorporating the photoswitch moiety to the \mbox{$\alpha$3-helix} of the PDZ3 domain, we were able to  allosterically alter its affinity towards the peptide ligand.\cite{Bozovic2020b}
Setting proteins in motion by photoswitching, we followed the response of the protein and its ligand in both UV/Vis and infrared spectral regions. Transient measurements exposed that the allosteric signal is transduced in a sequence of events, starting from the point of the perturbation –the \mbox{$\alpha$3-helix}- through the protein to the binding groove in 200~ns. The 200~ns timescale represents the speed of the allosteric signal within an isolated protein domain.

A time-dependent evaluation of biological processes is necessary in order to reconcile the structure-function relationship of proteins.  While a static structural depiction gives us insight into the protein's function, a broader kinetic picture is needed to comprehend the true dynamical nature of proteins, and capture intermediates that may not be obvious in equilibrium. Revealing the timescales in which signals travel through biological systems will help understanding how changes in intrinsic structure and dynamics ultimately affect and alter their natural function.\\

\noindent\textbf{Materials and Methods:} Photocontrollable PDZ3 domain and KETWV peptide were produced as described previously.\cite{Bozovic2020b} The purity of all samples was confirmed with mass spectrometry. The samples were dialysed against 20~mM NaPi, 15~mM NaCl, pH~6.8 buffer, and lyophilised. The samples were then dissolved in D$_2$O and left overnight in order to allow for H-D exchange, and subsequently re-lyoplilised. Prior to measurement, samples were dissolved in D$_2$O in appropriate concentrations. The concentration of protein sample was determined by the visible absorption of the photoswitch molecule at 325 nm, assuming an extinction coefficient of $\epsilon$=10,000 M$^{-1}$cm$^{-1}$. The concentration of the peptide was determined by the tryptophan absorption 280~nm. The concentration of all samples was cross-confirmed with quantitative amino acid analysis.\cite{cohen2001}

The time-resolved experiments were performed using two electronically synchronized Ti:sapphire laser systems with a repetition rate of \mbox{2.5 kHz}, allowing for delay up to \mbox{42 $\mu$s}.\cite{Bredenbeck2004} The wavelength of the pump laser was tuned to \mbox{840 nm} so that second harmonic generation in a BBO crystal produced pump pulses centered at a wavelength of \mbox{420 nm} to induce \textit{cis}-to-\textit{trans} isomerization. To minimize sample degradation, the compressing stage after light amplification was bypassed, resulting in stretched pulses of ca. \mbox{60 ps} duration FWHM and a power of \mbox{4.2 $\mu$J}. The beam diameter was \mbox{$\approx$ 135 $\mu$m} FWHM at the sample position. The pump pulses were mechanically chopped at half the repetition rate of the laser setup. The mid-IR probe pulses were obtained in an optical parametric amplifier,\cite{Hamm2000} and centered at \mbox{1640 cm$^{-1}$} with a pulse duration of \mbox{$\approx$ 100 fs} and a beam diameter of \mbox{$\approx$ 115 $\mu$m}. After the sample, they were passed through a spectrograph and detected in a \mbox{2$\times$64} MCT array detector with a spectral resolution of \mbox{$\approx$ 2 cm$^{-1}$/pixel}. The water vapor lines of the unpurged setup were used for  frequency calibration.  The polarization of the pump pulses was set to magic angle (\mbox{54.7\degree}) relative to the probe pulses and multichannel referencing as described in Ref.~\citenum{Feng2017} was used for noise suppression.

The visible probe experiments were performed as previously described.\cite{Buhrke2020} The probe pulses were obtained by focusing \mbox{$\approx$ 1 $\mu$J} of the \mbox{800 nm} probe laser into a \mbox{CaF\textsubscript{2}} window, which was continuously translated.

The samples (\mbox{$\approx$ 700 $\mu$L}; \mbox{1.1 mM} protein and \mbox{4.4 mM} peptide) were cycled through a closed system with a sample reservoir, a peristaltic pump and a flow-cell, all under inert atmosphere (\mbox{N\textsubscript{2}}). The reservoir was illuminated with a continuous wave laser at \mbox{370 nm} (\mbox{150 mW}, CrystaLaser) to prepare the sample in the \textit{cis}-state. The sample cell consisted of two \mbox{CaF\textsubscript{2}} windows separated by a \mbox{50 $\mu$m} Teflon spacer. The flow speed of the peristaltic pump was adjusted  to ensure sufficient sample exchange before the subsequent laser shot but to minimize flowing out of the sample at the latest time delay (\mbox{$\approx$ 42 $\mu$s}). Static reference spectra were acquired in an UV/Vis (Shimadzu UV-2450) and an FTIR (Bruker Tensor 27, resolution 2~cm$^{-1}$, N$_2$-cooled MCT detector) spectrometer, respectively, with sample parameters that were comparable to those in the time-resolved experiments.

To calculate the noise level of double-difference spectra, the data of the isolated protein, as well as those of the protein-peptide complex, were split into two individual sets with equal number of scans and an internal double difference was taken. The resulting spectrum was divided by a factor $\sqrt{2}$ to account for the reduction of scans. The results of this analysis are shown in Fig.~\ref{IR}c as dashed lines.\\

\noindent\textbf{Acknowledgement:} We thank Gerhard Stock and his group for many insightful discussions, the Functional Genomics Center Zurich, especially Serge Chesnov and Birgit Roth, for their work on the mass spectrometry and amino-acid analysis, and Livia Marie M\"uller for her contribution at an early stage of the project. The work has been supported by the Swiss National Science Foundation (SNF) through the NCCR MUST and Grant 200020B\_188694/1.\\


\makeatletter
\def\@biblabel#1{(#1)}
\makeatother

\def\bibsection{\section*{}} 

\noindent\textbf{References:}
\vspace{-1.5cm}


\providecommand{\latin}[1]{#1}
\makeatletter
\providecommand{\doi}
  {\begingroup\let\do\@makeother\dospecials
  \catcode`\{=1 \catcode`\}=2 \doi@aux}
\providecommand{\doi@aux}[1]{\endgroup\texttt{#1}}
\makeatother
\providecommand*\mcitethebibliography{\thebibliography}
\csname @ifundefined\endcsname{endmcitethebibliography}
  {\let\endmcitethebibliography\endthebibliography}{}

\end{document}